\documentclass[12pt,oneside, a4paper]{article}

\pdfoutput=1  
 
\ifx\pdfoutput\undefined
\usepackage[dvips,bookmarks=false]{hyperref}	
\else
\usepackage{hyperref}
\fi

\usepackage{orcidlink}
\usepackage{xcolor}
\definecolor{linkcolor}{rgb}{0.0, 0.47, 0.75}
\definecolor{citecolor}{rgb}{1.0, 0.5, 0.0}
\hypersetup{
  linkcolor  = linkcolor,
  citecolor  = linkcolor,
  urlcolor   = linkcolor,
  colorlinks = true
}

\usepackage{graphicx}
\usepackage{amssymb}
\usepackage{cite}
\usepackage{bm}
\usepackage{amsmath,amsthm}
\usepackage{cleveref}
\usepackage{siunitx}
\usepackage{slashed}
\numberwithin{equation}{section}
\usepackage[normalem]{ulem}

\newcommand{\capdef}{}
\newcommand{\mycaption}[2][\capdef]{\renewcommand{\capdef}{#2}%
       \caption[#1]{{\footnotesize #2}}}

\newcommand{\E}{\mathbb{E}}
\newcommand{\Var}{\operatorname{Var}}
\newcommand{\Cov}{\operatorname{Cov}}

\newcommand{\Corr}{\operatorname{Corr}}

\newcommand{\PGst}{\text{PG$^*$}}
\newcommand{\PGpl}{\text{PG$^+$}}
\newcommand{\cPGst}{\tilde\chi^2_\PGst}
\newcommand{\cPGpl}{\tilde\chi^2_\PGpl}

\begin{document}

\begin{titlepage}

\flushright{IFT-UAM/CSIC-26-135}

\begin{center}
\vspace*{2cm}
        {\Large\bf Testing the statistical compatibility of data sets with a common prior}
\vspace{1cm}

\renewcommand{\thefootnote}{\fnsymbol{footnote}}

{\bf Pilar Coloma}~\orcidlink{0000-0002-1164-9900}$^a$\footnote[1]{pilar.coloma@ift.csic.es},
{\bf Michele Maltoni}~\orcidlink{0000-0001-7444-4542}$^a$\footnote[2]{michele.maltoni@csic.es},
{\bf Thomas Schwetz}~\orcidlink{0000-0001-7091-1764}$^b$\footnote[3]{schwetz@kit.edu}
\vspace{5mm}

{\it%
$^a$ Instituto de F\'isica Te\'orica UAM-CSIC, C/Nicol\'as Cabrera 13-15, Campus de Cantoblanco, 28049 Madrid, Spain \\
$^b$  {Institut f\"ur Astroteilchenphysik, Karlsruher Institut f\"ur Technologie (KIT),\\
    Hermann-von-Helmholtz-Platz 1, 76344 Eggenstein-Leopoldshafen, Germany}}

\today
  
\vspace{8mm} 

\abstract{The \emph{Parameter Goodness-of-fit} (PG) test \cite{Maltoni:2003cu} offers a simple test statistic to study the compatibility of different data sets within a given theoretical model. In its original formulation it assumes that the data sets are statistically independent. We propose an extension of the PG test for the situation when some parameters of the model are constrained by an auxiliary data set called prior here. As the prior is imposed for all data sets, they are no longer statistically independent. We show that under the Gaussian approximation, the prior-constrained PG test statistic follows a generalised $\chi^2$ distribution, which can be used to obtain $p$-values in a straight-forward way. In the approximation of a strong prior constraint, the test converges towards the standard PG test for the unconstrained model parameters.}

\end{center}

\end{titlepage}

\renewcommand{\thefootnote}{\arabic{footnote}}
\setcounter{footnote}{0}

\tableofcontents

\section{Introduction}

A common statistics problem in fundamental physics is to test, whether different data sets are compatible within a given theoretical model. In Ref.~\cite{Maltoni:2003cu} a frequentist test has been discussed, called the \emph{Parameter Goodness-of-fit} (PG) test, originally introduced in the context of sterile neutrino searches~\cite{Maltoni:2002xd}. Various compatibility tests or tension metrics are applied in the cosmological context (mostly in a Bayesian framework), see Refs.~\cite{DES:2020hen,Leizerovich:2023qqt} for reviews or \cite{Amendola:2012wc,Raveri:2018wln,Lin:2019zdn,Gariazzo:2023joe,Mello:2024tor} for other examples for cosmological applications.
In this work we focus on the PG test in a frequentist framework. This test is based on simple $\chi^2$ variables and under rather general conditions -- similar to the ones required for Wilks' theorem~\cite{Wilks:1938dza} to apply -- the corresponding test statistic follows a $\chi^2$ distribution with the number of degrees of freedom determined by the number of common parameters of the model used to describe the data sets to be tested for compatibility~\cite{Maltoni:2003cu}.

A requirement of the applicability of this analytic result for the PG test is the statistical independence of the tested data sets. However in many situations of interest the model used to describe the data may depend on nuisance parameters, which are constrained by a different set of observations or external knowledge  called ``prior'' in the following. The important observation is that in the presence of a common prior the data sets are no longer statistically independent, as in general the common prior leads to correlations. Therefore the standard PG test cannot be applied in this case. Two typical examples are: $(i)$ The data sets may depend on common systematic uncertainties, which are parametrised by nuisance parameters subject to external measurements. For example, the data sets could correspond to different neutrino oscillation experiments of the same type (\textit{e.g.}, reactor experiments, or accelerator neutrino experiments), suffering from the same neutrino flux or detection cross section uncertainties. $(ii)$ In some applications one may want to impose an external constraint to one or more of the model parameters, and ask the question, whether the data sets of interest are compatible under the external constraint. 

The aim of this work is to generalise the PG test and allow for the presence of a common prior. We will introduce two versions of a modified test statistic, which allow to test the compatibility of data sets linked by a common prior. 
In general, the prior-constrained PG statistics will not follow a simple $\chi^2$ distribution. However, we will show that under the usual Gaussian approximation the modified statistics will be distributed as a certain linear combination of $\chi^2$ variables, which is a particular case of a so-called generalised $\chi^2$ distribution \cite{wiki}. While no closed form exists for the generalised $\chi^2$ distribution, it has a simple analytic characteristic function, which by numerical integration allows to construct the probability distribution function and the cumulative distribution function of the test statistic. Along with this paper we provide a Python script to calculate the relevant distributions \cite{ref-code}. 
Furthermore, we can use our analytical results to define the conditions under which we expect the standard PG test to apply approximately (namely in the limit of ``strong'' prior, to be defined in detail below). 

In the context of cosmology, a test similar to the PG is often called DMAP (Difference of log-likelihoods at their Maximum a Posteriori point) \cite{Raveri:2018wln}. Ref.~\cite{Raveri:2019gdp} discusses tension metrics in the case of correlated data sets, which has some overlap with our considerations. In particular, they discuss the DMAP test and find that under certain approximations its distribution is a linear combination of $\chi^2$ variables, in agreement with our findings. Our work differs from Ref.~\cite{Raveri:2019gdp} in that we provide two variants of possible test statistics (one of them being strictly positive), formulate all correlations explicitly in terms of the prior data set, we work in the pure frequentist $\chi^2$ framework, and we provide simple-to-use recipes for the $p$-value calculation in the Gaussian limit, based on the characteristic function of the test statistic distribution.  

The outline of the rest of the paper is as follows. In \cref{sec:setup} we set the notation and define two versions of a test statistics for the prior-constrained parameter goodness-of-fit, denoted by \PGst\ and \PGpl.  
Their analytic distributions under the Gaussian approximation are derived in \cref{sec:PGst,sec:PGpl}, respectively. In \cref{sec:numres} we consider a simple toy experiment simulation inspired by neutrino oscillation disappearance measurements, which we use to study the \PGst\ and \PGpl\ test statistics by Monte Carlo simulations. We verify that the Monte Carlo distribution agrees with the analytic results in the Gaussian approximation. Furthermore, we provide a comparison of the two test statistics and further discussion of their properties, as well as an illustration of the power of the test to reject a model with tension between data sets (including a prior). 
We summarize our results in \cref{sec:summary}. Throughout the main text we limit the discussion to the compatibility of two data sets plus a prior, while we comment on the multi-data set case in appendix~\ref{appendix}. 

\section{Setup and definition of the test statistics}
\label{sec:setup}

Let us consider two statistically independent data sets $A$ and $B$, consisting of $N_A$, $N_B$ data points respectively. 
We want to describe these data with a model depending on $p$ parameters $\bm{\theta}$ common for both data sets. We can construct $\chi^2$ functions as
\begin{align}
  \chi^2_A(\bm{\theta}) &= \chi^2_{A,{\rm min}} + \Delta\chi^2_A(\bm{\theta}) \,,
  \\
  \chi^2_B(\bm{\theta}) &= \chi^2_{B,{\rm min}} + \Delta\chi^2_B(\bm{\theta}) \,, 
\end{align}
where $\chi^2_{S,{\rm min}} = \min_{\bm{\theta}}\chi^2_S(\bm{\theta})$ for $S=A,B$.
The parameter goodness-of-fit test \cite{Maltoni:2003cu} provides a test-statistic for testing the compatibility of the two data sets under the hypothesis that the model is true:
\begin{align}
  \chi^2_{\rm PG} &=
\min_{\bm{\theta}}
\left[
\chi_A^2(\bm{\theta})
+
\chi_B^2(\bm{\theta})
\right]
- \chi^2_{A,{\rm min}} - \chi^2_{B,{\rm min}} \label{eq:PG1}
\\
&=
\left[
\chi_A^2(\widehat{\bm{\theta}}_{AB}) - \chi^2_{A,{\rm min}}
\right] + \left[
\chi_B^2(\widehat{\bm{\theta}}_{AB}) - \chi^2_{B,{\rm min}}
\right]
\label{eq:PG2}
\\
&=\Delta\chi^2_A(\widehat{\bm{\theta}}_{AB}) + \Delta\chi^2_B(\widehat{\bm{\theta}}_{AB}) 
 \,, \label{eq:PG3}
\end{align}
where $\widehat{\bm{\theta}}_{AB}$ denotes the best fit parameters from the combination of both data sets.
Under general conditions, similar to the ones required for Wilks' theorem to hold, $\chi^2_{\rm PG}$ will be $\chi^2_p$ distributed, \textit{i.e.}, a $\chi^2$ distribution with $p$ degrees-of-freedom \cite{Maltoni:2003cu}. We will call this limit in the following ``Gaussian approximation''.

Now we want to consider the situation where, in addition to the $p$ parameters $\bm{\theta}$, the model depends also on $q$ nuisance parameters $\bm{\xi}$, which are constrained by another independent data set called ``prior''. We describe the prior by
\begin{equation}
  \chi^2_P(\bm{\xi}) = \chi^2_{P,{\rm min}} 
  + \Delta\chi^2_P(\bm{\xi}) \,. 
\end{equation}
We want to ask now the question, whether $A$ and $B$ are consistent under the additional constrain imposed by the prior data set on the nuisance parameters. The nuisance parameters can describe common systematic uncertainties, or an external constraint on some of the model parameters (in which case we will replace $p \to p-q$ so that the total number of parameters remains the same, and that $p$ consistently counts the parameters \emph{not} affected by the prior).

We define individual $\chi^2$ functions for the two data sets including the prior as
\begin{align}
  \chi^2_{AP}(\bm{\theta},\bm{\xi}) &= \chi^2_A(\bm{\theta},\bm{\xi}) + \Delta\chi^2_P(\bm{\xi})\,, \label{eq:chisqAP}\\
  \chi^2_{BP}(\bm{\theta},\bm{\xi}) &= \chi^2_B(\bm{\theta},\bm{\xi}) + \Delta\chi^2_P(\bm{\xi})\,, \label{eq:chisqBP} 
\end{align}
where we ignore $\chi^2_{P,{\rm min}}$ and add only $\Delta\chi^2_P(\bm{\xi})$, as we are only interested in the constraint on the nuisance parameters imposed by the prior. Note, that now $\chi^2_{AP}$ and $\chi^2_{BP}$ are no longer statistically independent, because they include the same statistical fluctuations from the prior data set. Therefore, a straight-forward application of the original PG test is not possible, as it applies only for statistically independent data sets. 

We are proposing in the following two possible test statistics including the prior constraint, which both are natural generalizations of the PG test statistic. First we define the \PGst\ test statistic by 
\begin{align}
  \cPGst &=
\min_{\bm{\theta},\bm{\xi}}
\left[
\chi_A^2(\bm{\theta},\bm{\xi})
+
\chi_B^2(\bm{\theta},\bm{\xi})
+
\Delta\chi_P^2(\bm{\xi})
\right]
- \chi^2_{AP,{\rm min}} - \chi^2_{BP,{\rm min}} \,,\label{eq:PGst1}
\end{align}
generalizing the PG version displayed in \cref{eq:PG1}.
Unlike the original $\chi^2_{\rm PG}$, this quantity is not strictly positive and can also acquire negative values. Hence, it cannot follow a $\chi^2_\nu$ distribution. Indeed, as we are going to show below, in the Gaussian approximation $\cPGst$ will follow a generalised $\chi^2$ distribution. We use the tilde to indicate that this quantity in general is not expected to be $\chi^2$ distributed.

As an alternative version, we define the \PGpl\ test statistic, inspired by \cref{eq:PG2}:
\begin{equation}
\cPGpl =
\Big[ 
\chi_{AP}^2(\widehat{\bm{x}}_{ABP}) - \chi^2_{AP,{\rm min}}
\Big] + \Big[ 
\chi_{BP}^2(\widehat{\bm{x}}_{ABP}) - \chi^2_{BP,{\rm min}} 
\Big] \,,
\label{eq:PGpl}
\end{equation}
where $\bm{x} = ( \bm{\theta},\bm{\xi})$ and $\widehat{\bm{x}}_{ABP}$ denotes the best fit parameters from the combination of $A,B$ and the prior $P$. Each bracket of \cref{eq:PGpl} is non-negative and therefore $\cPGpl \ge 0$ by construction, therefore the `+' symbol in the acronym. As we will show, its distribution will also be a generalised $\chi^2$ distribution in the Gaussian limit.

Useful identities are 
\begin{align}
  \cPGst &= \chi^2_{{\rm PG},ABP} - \chi^2_{{\rm PG},AP} - \chi^2_{{\rm PG},BP} \,, \label{eq:PGst2} \\
  \cPGpl &= \cPGst + \Delta\chi^2_P(\widehat\xi_{ABP}) \,. \label{eq:PGpl2}
\end{align}
Here, $\chi^2_{{\rm PG},ABP}$ denotes the standard PG test statistic for the consistency of the three independent data sets $A$, $B$, $P$ defined by
\begin{equation}
  \chi^2_{{\rm PG},ABP} = 
\min_{\bm{\theta},\bm{\xi}}
\left[
\chi_A^2(\bm{\theta},\bm{\xi})
+
\chi_B^2(\bm{\theta},\bm{\xi})
+
\Delta\chi_P^2(\bm{\xi})
\right] - \chi^2_{A,{\rm min}} - \chi^2_{B,{\rm min}} \,,
\end{equation}
which follows a $\chi^2_\nu$ distribution with $\nu = 2(p+q) + q - (p+q) = p + 2q$ degrees of freedom \cite{Maltoni:2003cu},
and $\chi^2_{{\rm PG},AP}, \chi^2_{{\rm PG},BP}$ are the PG test statistics for the consistency of the data sets A and B with the prior, respectively. Hence, the three terms in 
\cref{eq:PGst2} are individually distributed as $\chi^2_{p+2q}, \chi^2_{q}, \chi^2_{q}$, respectively. However, it is important to note that they are not statistically independent, as they are constructed from different combinations of the same random data.  

\Cref{eq:PGpl2} shows that $\cPGst$ and $\cPGpl$ differ just by a prior term. This term appears only once in $\cPGst$ but twice in $\cPGpl$, ensuring positivity. As a consequence, \PGpl\ gives somewhat more weight to a tension appearing in the prior, whereas \PGst\ has more weight on the ``data'' part of the combination. In the following we will discuss the \PGst\ and \PGpl\ test statistics in turn.

\section{The \PGst\ test statistic} \label{sec:PGst}

\subsection{Derivation of the distribution for $p=1$, $q=1$}
\label{sec:one-param}

We first provide a derivation of the distribution of $\cPGst$ for the simplest case of $p=q=1$, \textit{i.e.}, one single model parameter $\theta$ and one single nuisance parameter $\xi$. The derivation makes use of standard methods from the theory of quadratic forms, see \textit{e.g.}, \cite{quadratic-forms} for a text book.

In the Gaussian approximation, the $\chi^2$ functions can be written in the following way:
\begin{align}
\chi_A^2(\bm{x})
&=
\chi^2_{A,{\rm min}} +
(\bm{x}-\widehat{\bm{x}}_A)^T
V_A^{-1}
(\bm{x}-\widehat{\bm{x}}_A) \,, \label{eq:chisqA}
\\
\chi_B^2(\bm{x})
&=
\chi^2_{B,{\rm min}} +
(\bm{x}-\widehat{\bm{x}}_B)^T
V_B^{-1}
(\bm{x}-\widehat{\bm{x}}_B)\,,\label{eq:chisqB}
\\
\chi_P^2(\xi)
&=
\chi^2_{P,{\rm min}} +\frac{(\xi-\widehat{\xi}_P)^2}{s_P} \label{eq:chisqP}\,.
\end{align}
Here, the hat denotes the best-fit parameters, \textit{i.e.}, the parameter values at the $\chi^2$ minimum of the corresponding data set, while $V_A$ and $ V_B$ are the covariance matrices of $\bm{x}$ from the data sets $A$ and $B$, respectively. They are either predicted by the model, or estimated by
\begin{equation}
  (V_S^{-1})_{xy} = \frac{1}{2} \left. \frac{\partial^2\chi^2_S}{\partial x\partial y} \right|_{\widehat{\bm{x}}}
\end{equation}
with $x,y\in\{\theta,\xi\}$ and $S=A,B$. Similarly, $s_P$ is the variance of $\xi$ from the prior. The expressions in \cref{eq:chisqA,eq:chisqB,eq:chisqP} hold exactly for a model with linear parameter dependence; in general, they are valid sufficiently close to the $\chi^2$ minima, as long as these are located far enough from the boundaries of the parameter space. 

Let us consider the determination of the nuisance parameter $\xi$ from the data sets $A$ and $B$ without the prior. The marginal variances for $\xi$ from data set $A,B$ are given by the $\xi\xi$ element of the corresponding covariance matrix:
\begin{equation} \label{eq:sAB}
  s_A = (V_A)_{\xi\xi}\,,
  \qquad
  s_B = (V_B)_{\xi\xi} \,. 
\end{equation}
We can now evaluate the consistency of the data sets $A$ and $B$ with the prior by considering the standardised discrepancies
\begin{equation}\label{eq:disc-AB}
  d_{AP} = \frac{\widehat{\xi}_A - \widehat{\xi}_P}{\sqrt{s_A + s_P}} \,,\qquad 
  d_{BP} = \frac{\widehat{\xi}_B - \widehat{\xi}_P}{\sqrt{s_B + s_P}} \,.
\end{equation}
Under the null-hypothesis of the model being true, they will be standard normal distributed. Furthermore, one can show that $\chi^2_{{\rm PG},SP} = d_{SP}^2$ $(S=A,B)$, and therefore \cref{eq:PGst2} can be written as 
\begin{equation}\label{eq:PGst3}
  \cPGst = \chi^2_{{\rm PG},ABP} - d_{AP}^2 - d_{BP}^2 \,.  
\end{equation}
The three terms are individually distributed as $\chi^2_3,\chi^2_1,\chi^2_1$ but, as emphasized above, they are not statistically independent.

In order to write $\cPGst$ in terms of statistically independent quantities, we use the correlation of the discrepancies $d_{AP}$ and $d_{BP}$.  The
estimators from $A$, $B$, and $P$ are independent, hence we have
\begin{equation}
\Cov(
\widehat{\xi}_A-\widehat{\xi}_P,\,
\widehat{\xi}_B-\widehat{\xi}_P
)
= \Var(\widehat{\xi}_P) =s_P \,,
\end{equation}
and the correlation between $d_{AP}$ and $d_{BP}$ is given by 
\begin{equation}
\rho_{AB}
\equiv
\operatorname{Corr}(d_{AP},\, d_{BP})
=
\frac{s_P}{
\sqrt{(s_A+s_P)(s_B+s_P)}
} \,.
\label{eq:rho}
\end{equation}
Since $s_A,s_B,s_P>0$, we have $0<\rho_{AB}<1$. The limiting values $0$ and $1$ are obtained for an infinitely strong ($s_P \to 0$) or infinitely weak ($s_P \to \infty$) constraint from the prior $P$, respectively.

We are now in the position to derive the distribution of $\cPGst$. Considering the form in \cref{eq:PGst3}, we note that the first term, $\chi^2_{{\rm PG},ABC}$, being distributed as $\chi^2_3$, is the squared norm of a three-dimensional vector of standard-normal random variables:
\begin{equation}
  \chi^2_{{\rm PG},ABC} = \bm{g}^2 \,, \qquad \bm{g}\sim N(\bm{0},I_3) \,.
\end{equation}
Now we can take the $\bm{g}$ as basis to express the discrepancies $d_{AP}$ and $d_{BP}$ providing the remaining terms in \cref{eq:PGst3} by expressing 
$d_{AP}^2$ and $d_{BP}^2$ as rank-one projection quadratic
forms. Taking into account the correlation \cref{eq:rho}, we write \cref{eq:PGst3} as
\begin{equation}\label{eq:K}
  \cPGst = \bm{g}^TK\bm{g} \,,
\end{equation}
where
\begin{equation}
K=I_3-\bm{a}\bm{a}^T-\bm{b}\bm{b}^T \,,  
\end{equation}
and $\bm{a}$ and $\bm{b}$ are unit vectors satisfying
\begin{equation}\label{eq:ab}
\bm{a}^T\bm{b}=\rho_{AB} \,.  
\end{equation}
The matrix
$\bm{a}\bm{a}^T+\bm{b}\bm{b}^T$
has eigenvalues
$(1+\rho_{AB},1-\rho_{AB},0)$, which implies that 
$K$ has eigenvalues
$(1, \rho_{AB}, -\rho_{AB})$. Therefore, upon diagonalizing \cref{eq:K} we obtain for the distribution of the test statistic:
\begin{equation}
\cPGst
\overset{d}{=}
G_0^2+\rho_{AB} (G_1^2- G_2^2) \,,
\label{eq:PGstG}
\end{equation}
where $G_0,G_1,G_2$ are three independent standard-normal random variables. Or equivalently,
\begin{equation}
\cPGst
\overset{d}{=}
X_0+\rho_{AB} (X_1 -  X_2) \,,
\label{eq:PGstX}
\end{equation}
where $X_i \sim \chi^2_1$ are mutually independent $\chi^2$ variables with 1 degree of freedom.
Thus, $\cPGst$ has a generalized $\chi^2$ distribution \cite{wiki}, see also \cite{imhof:1961}. Its
distribution depends on the experimental setup only through the
single dimensionless parameter $\rho_{AB}$, \cref{eq:rho}.
For $0<\rho_{AB}<1$, the negative eigenvalue $-\rho_{AB}$ implies that
$\cPGst$ can take negative values. Hence it is not, in general, an
ordinary $\chi^2$ random variable. From \cref{eq:PGstX} we see that when taking the expectation value of $\cPGst$, the two terms proportional to $\rho_{AB}$ cancel and we obtain $\E\left[\cPGst\right] = \E\left[\chi^2_1\right] = 1$.

Using the independent representation in \cref{eq:PGstX}, the characteristic function $\phi_z(t) \equiv \mathbb{E}[e^{itz}]$ of
$z =\cPGst$ is  \cite{wiki,imhof:1961}
\begin{equation}
\phi_z(t)
=
(1-2it)^{-1/2}
(1-2it\rho_{AB})^{-1/2}
(1+2it\rho_{AB})^{-1/2}
=
\frac{1}{\sqrt{(1-2it)(1+4\rho_{AB}^2t^2)}} \,.
\label{eq:Zprime-CF}
\end{equation}
This characteristic function completely specifies the null
distribution of the test statistic; the probability distribution function (PDF) $f(z)$ and the cumulative distribution function (CDF) $F(z) = \int^z dz'f(z')$ can be calculated in a straight-forward way:
\begin{align}
  f(z) &= \frac{1}{2\pi} \int dt \, e^{-izt} \phi_z(t) \,, \label{eq:pdf}
  \\
  F(z) &= \frac{1}{2} - 
  \frac{1}{\pi} \int_0^\infty dt \, \frac{1}{t} \, {\rm Im} \left[e^{-izt} \phi_z(t) \right] \, \label{eq:cdf}
\end{align}
where for the latter we have used the Gil-Pelaez theorem. This expression allows directly to calculate $p$-values for the \PGst\ test.

\paragraph{Limiting cases: weak vs strong priors.}

From \cref{eq:PGstX} it is clear that, for small $\rho_{AB}$, $\cPGst$ will be approximately distributed as $\chi^2_1$ (which holds exactly for $\rho_{AB}=0$). Note that for $\rho_{AB} \ll 1$ it is enough that either $s_A$ or $s_B$ is much larger than $s_P$, see \cref{eq:rho}. Hence, if only one of the data sets provides a much weaker constraint on $\xi$ than the prior we expect a distribution close to $\chi^2_1$. This will be referred to as the ``strong prior'' limit.

Conversely, in the ``weak prior'' limit $s_P \gg s_A,s_B$, and $\rho_{AB} \approx 1$. In this case, \cref{eq:PGstX} implies that $\cPGst \sim \chi^2_2 - \chi^2_1$, which does not converge towards a normal $\chi^2$ but still remains a generalized $\chi^2$ distribution. Hence, the limit $s_P\to \infty$ is \emph{not} equivalent to removing the prior completely (in which case we would recover the normal PG, distributed as $\chi^2_2$). The reason is that although the limit $s_P\to\infty$ removes effectively any constraint on $\xi$ from the prior, the best fit value $\widehat{\xi}_P$ will obtain fluctuations of order $\sqrt{s_P}$, tending to infinity, with the effect that the last term in \cref{eq:chisqP} will still contribute as $\chi^2_1$, irrespective of the value of $s_P$. 

\subsection{Generalization to an arbitrary number of parameters}
\label{sec:multi-param}

The above reasoning can be generalised to any number of model parameters $p>1$ and nuisance parameters $q>1$. Without providing the full derivation we sketch the main idea below. 
The expressions for $\chi^2_A$ and $\chi^2_B$ from \cref{eq:chisqA,eq:chisqB} generalize with $\bm{x} = (\bm{\theta},\bm{\xi})$ being a $(p+q)$-vector. In generalisation of \cref{eq:sAB} we obtain the $q\times q$ covariance matrices $S_A$, $S_B$ for the nuisance parameters from data sets $A,B$ as the $\bm{\xi}$ sub-block of the covariance matrices $V_A$ and $V_B$, whereas $S_P$ denotes now the prior covariance matrix for $\bm{\xi}$. We define the two discrepancy vectors
\begin{equation}
\bm{D}_{AP}
=
\widehat{\bm{\xi}}_A-\widehat{\bm{\xi}}_P \,,
\qquad
\bm{D}_{BP}
=
\widehat{\bm{\xi}}_B-\widehat{\bm{\xi}}_P \,.
\end{equation}
Under the common-parameter null hypothesis, their covariance matrices are
\begin{align}
\Cov(\bm{D}_{AP})
&=
S_A+S_P,\\
\Cov(\bm{D}_{BP})
&=
S_B+S_P,\\
\Cov(\bm{D}_{AP},\bm{D}_{BP})
&=
S_P.
\end{align}
Therefore, we can define standardised discrepancies as
\begin{align}
\bm{d}_{AP}
&=
(S_A+S_P)^{-1/2}\bm{D}_{AP},\\
\bm{d}_{BP}
&=
(S_B+S_P)^{-1/2}\bm{D}_{BP}.
\end{align}
which are vectors of standard normal variables with the correlation matrix
\begin{equation}
R \equiv \Corr (\bm{d}_{AP}, \bm{d}_{BP}) = 
(S_A+S_P)^{-1/2}
S_P
(S_B+S_P)^{-1/2} \,.
\label{eq:R}
\end{equation}
Let $\rho_1,\ldots,\rho_q$ be the
singular values of $R$. These are the canonical correlations between
the two standardized discrepancy vectors $\bm{d}_{AP}$ and
$\bm{d}_{BP}$. Equivalently, the squared canonical correlations
$\rho_i^2$ are the eigenvalues of
\begin{equation}
RR^T
=
(S_A+S_P)^{-1/2}
S_P
(S_B+S_P)^{-1}
S_P
(S_A+S_P)^{-1/2}
\end{equation}
and they satisfy $0\leq \rho_i\leq 1$. For $q=1$, \cref{eq:R} reduces to \cref{eq:rho}.
Now we can write \cref{eq:PGst2} as
\begin{equation}
  \cPGst = \chi^2_{{\rm PG},ABP} - \bm{d}_{AP}^T\bm{d}_{AP} - \bm{d}_{BP}^T\bm{d}_{BP} \,,  
\end{equation}
with the three terms being $\chi^2_{p+2q},\chi^2_q,\chi^2_q$. Similar as above, we consider
a standard Gaussian
vector $\bm{g}$ of dimension $p+2q$. Within the space of $\bm{g}$, the $AP$ and $BP$ terms in the above sum are rank-$q$ orthogonal projection quadratic forms. Consequently, we can write
\begin{equation}
\cPGst = \bm{g}^TK\bm{g} \,,
\qquad
K=I-P_A-P_B \,,
\end{equation}
where $P_A$ and $P_B$ are rank-$q$ orthogonal projectors
corresponding to the $A-C$ and $B-C$ discrepancy subspaces.
Geometrically, the cosines of the principal angles between these two subspaces are
the canonical correlations
$\rho_1,\ldots,\rho_q$, in generalization of \cref{eq:ab}. The eigenvalues of $K$ are
\begin{equation}
\underbrace{1,\ldots,1}_{p\ \text{times}},
\quad
\rho_1,-\rho_1,
\quad\ldots,\quad
\rho_q,-\rho_q \,.
\label{eq:eigenvalues}
\end{equation}
Therefore, under the common-parameter null hypothesis,
\begin{equation}
\cPGst
\overset{d}{=}
\sum_{j=1}^{p}G_j^2
+
\sum_{i=1}^{q}
\rho_i
\left(
U_i^2-V_i^2
\right),
\label{eq:Zprime-normal}
\end{equation}
where $G_j$, $U_i$, $V_i$ are mutually independent standard-normal variables. Or equivalently,
\begin{equation}
\cPGst
\overset{d}{=}
\chi_p^2
+
\sum_{i=1}^{q}
\rho_i \left(X_i - Y_i\right)\,,
\label{eq:Zprime-chi-2}
\end{equation}
with $X_i,Y_i \sim \chi^2_1$ and all $\chi^2$ variables mutually
independent.
Thus we again obtain that $\cPGst$ is a generalized $\chi^2$ random
variable. It is an ordinary $\chi^2$ variable only when all canonical correlations vanish.

We see from \cref{eq:Zprime-chi-2} that
\begin{equation}\label{eq:expPGst}
    \E\left[\cPGst\right] = \E\left[\chi^2_p\right] = p \,,
\end{equation}
independent of the prior. Hence, a value of $\cPGst \gg p$ signals a low $p$-value, irrespective of the prior.

From Eq.~\eqref{eq:Zprime-chi-2}, the characteristic function is \cite{wiki}
\begin{align}
\phi_{z}(t)
&=
(1-2it)^{-p/2}
\prod_{i=1}^{q}
(1-2it\rho_i)^{-1/2}
(1+2it\rho_i)^{-1/2}\\
&=
(1-2it)^{-p/2}
\prod_{i=1}^{q}
\left(
1+4\rho_i^2t^2
\right)^{-1/2} \\
&=
(1-2it)^{-p/2}
\det\left(
I_q+4t^2RR^T
\right)^{-1/2} \,.
\label{eq:characteristic-function-matrix}
\end{align}

Hence, the task in the general case is to find the $\bm{\xi}$ correlation matrices $S_A$, $S_B$, $S_P$, calculate the matrix $R$ and the canonical correlations $\rho_i$. Once they are known, we can obtain the PDF or CDF via \cref{eq:pdf,eq:cdf} using the above expression for the characteristic function. Similar as for the one-parameter case, we can define the limit of ``strong prior'' by $\rho_i\ll 1$ ($i=1,\ldots q$), in which case the \PGst\ becomes the standard PG, distributed as a normal $\chi^2$ variable with $p$ degrees of freedom.

\section{The \PGpl\ test statistic}\label{sec:PGpl}

Let us move now to the \PGpl\ test and derive its distribution under the Gaussian approximation. We work this out here only for the case $p=q=1$ (with $p>1$ and $q=1$ being a trivial extension); we comment on the general case at the end of this section. According to \cref{eq:PGpl2} we have
\begin{equation}\label{eq:PGpl1}
  \cPGpl = \cPGst + \frac{(\widehat\xi_{ABP} - \widehat\xi_P)^2}{s_P} \,.
\end{equation}
We proceed in analogy to \cref{sec:one-param}. First we need the distribution of the last term in \cref{eq:PGpl1}. We start by calculating $s_{ABP}$, the variance of $\widehat\xi_{ABP}$. Define
$F = V^{-1}_A + V^{-1}_B$ and
\begin{equation}
  F_{P} = F + 
  \begin{pmatrix}
    0&0 \\ 0& 1/s_P
  \end{pmatrix}.
\end{equation}
Then we have
\begin{equation}\label{eq:sABP}
  s_{ABP} = (F_{P}^{-1})_{\xi\xi} = \frac{1}{\frac{1}{s_P} + F_{\xi\xi} - \frac{F_{\theta\xi}^2}{F_{\theta\theta}}}  \,.
\end{equation}
Next we find
\begin{equation} \label{eq:cov}
  \Cov(\widehat\xi_{ABP},\widehat\xi_P) = 
  \Cov(\widehat\xi_{ABP},\widehat\xi_A) = 
  \Cov(\widehat\xi_{ABP},\widehat\xi_B) = s_{ABP} \,. 
\end{equation}
It follows that $\Var(\widehat\xi_{ABP} - \widehat\xi_P) = s_P - s_{ABP}$, and therefore
\begin{equation}
  d_P = \frac{\widehat\xi_{ABP} - \widehat\xi_P}{\sqrt{s_P - s_{ABP}}} \sim N(0,1) \,.
\end{equation}
We see that the last term in \cref{eq:PGpl1} is a re-scaled $\chi^2_1$:
\begin{equation}
  \frac{(\widehat\xi_{ABP} - \widehat\xi_P)^2}{s_P} = \kappa d_P^2 \sim \kappa \chi^2_1 \,,\qquad
  \kappa = \frac{s_P - s_{ABP}}{s_P} \,,
\end{equation}
with $0\le \kappa \le 1$, and we can write
\begin{equation}
  \cPGpl = \chi^2_{{\rm PG},ABP} - d^2_{AP} - d^2_{BP} + \kappa d_P^2 \,,
\end{equation}
where we have used the discrepancies $d_{SP}$ defined in \cref{eq:disc-AB}. With \cref{eq:cov} we have $\rho_{AB}$ defined as in \cref{eq:rho} and, additionally,
\begin{align}
\rho_{AP}
&\equiv
\operatorname{Corr}(d_{AP},d_{P})
=
\frac{s_P}{
\sqrt{(s_A+s_P)(s_P-s_{ABP})}
} \,,
\label{eq:rhoAP}\\
\rho_{BP}
&\equiv
\operatorname{Corr}(d_{BP},d_{P})
=
\frac{s_P}{
\sqrt{(s_B+s_P)(s_P-s_{ABP})}
} \,.
\label{eq:rhoBP}
\end{align}

In analogy to \cref{eq:K} we write
\begin{equation}\label{eq:K-PGpl}
  \cPGpl = \bm{g}^TK\bm{g} \,, \qquad \bm{g} \sim N(0,I_3) 
\end{equation}
and
\begin{equation}\label{eq:K-PGpl2}
K=I_3-\bm{a}\bm{a}^T-\bm{b}\bm{b}^T + \kappa \bm{c}\bm{c}^T\,,  
\end{equation}
with $\bm{a},\bm{b},\bm{c}$ are unit vectors satisfying
\begin{equation}
\bm{a}^T\bm{b}=\rho_{AB} \,,\qquad  
\bm{a}^T\bm{c}=\rho_{AP} \,,\qquad  
\bm{b}^T\bm{c}=\rho_{BP} \,.  
\end{equation}
Diagonalising $K$ leads to
\begin{equation}\label{eq:PGpl-X}
  \cPGpl = \sum_{i=1}^3 \lambda_i X_i^2 \,,
\end{equation}
where $\lambda_i$ are the eigenvalues of $K$ and $X_i$ are mutually independent standard normal variables. Hence, $\cPGpl$ also follows a generalised $\chi^2$ distribution with the characteristic function 
\begin{equation}
  \phi_z(t) = \prod_{i=1}^3 \left(1 - 2it \lambda_i \right)^{-1/2} \,.
  \label{eq:phiz-PGpl}
\end{equation}

The eigenvalues can be obtained from diagonalising the matrix $K$ numerically by choosing an explicit basis for the vectors $\bm{a},\bm{b},\bm{c}$ or as a solution of the cubic equation:\footnote{In the case of more model parameters $p>1$ and $q=1$, there will be $p-1$ eigenvalues equal to one and the three non-trivial eigenvalues given as a solution to the same cubic equation.}
\begin{equation}
  \lambda^3 - (1+\kappa)\lambda^2 
  + \left[ \kappa(\rho_{AP}^2 + \rho_{BP}^2) - \rho^2_{AB} \right] \lambda
  + \rho_{AB} \left[(1+\kappa)\rho_{AB} - 2\kappa\, \rho_{AP}\, \rho_{BP} \right] = 0 \,.
\end{equation}
It can be shown that all $\lambda_i \ge 0$, and therefore $\cPGpl$ is always non-negative, as per construction. 

\paragraph{Limiting cases: weak vs strong priors.} 

Let us consider again the strong and weak prior limits for this test statistic. In the strong prior limit ($s_P\to 0 $) we have $\kappa \to 0$ and all $\rho_{XY} \to 0$. The cubic equation becomes $\lambda^3 - \lambda^2 = 0$, which has the solutions $\lambda_1=\lambda_2=0$, $\lambda_3 = 1$ and therefore $\cPGpl \sim \chi^2_1$, as expected for the PG once $\xi$ is fixed. Conversely, in the weak prior limit ($s_P\to \infty $) we have $\kappa \to 1$ and all $\rho_{XY} \to 1$. The cubic equation in this case becomes $\lambda^3 - 2\lambda^2 + \lambda= 0$, which has the solutions $\lambda_1=0$, $\lambda_2 = \lambda_3 = 1$. This implies $\cPGpl \sim \chi^2_2$, which corresponds to the PG of $A$ versus $B$ without a prior. 
Hence, the \PGpl\ has the property that both the strong and weak prior limits converge towards a standard $\chi^2$ distribution.\footnote{Since in the \PGpl\ there is an equal number of positive and negative prior terms in the definition, all remnant prior contributions cancel in the limit $s_P\to\infty$, in contrast to the \PGst, where one unbalanced prior term remains.} 

From \cref{eq:PGpl-X} we also see that
\begin{equation}
    \E\left[\cPGpl\right] = \sum_{i=1}^3 \lambda_i = {\rm Tr}[K] = 1+\kappa \,,
\end{equation}
where we have used that each projector in \cref{eq:K-PGpl2} has trace 1. This is consistent with the above discussed limiting distributions for strong and weak prior, with $\kappa$ interpolating between $0\to 1$. For $p>1$, $q=1$ this generalises to 
$\E\left[\cPGpl\right] = p+\kappa$. 

\paragraph{Generalization to an arbitrary number of parameters.} 

Finally, let us briefly comment on the case $p>1$ and $q>1$. In principle, the distribution under the Gaussian approximation can be calculated using similar methods as in \cref{sec:multi-param}, which we do not discuss explicitly here (the corresponding distribution is provided numerically in the accompanying code~\cite{ref-code}).
Let us just remark that also in the general case, in the strong and weak prior limit the \PGpl\ statistics converges towards the standard PG. This follows from the definition in \cref{eq:PGpl} together with \cref{eq:chisqAP,eq:chisqBP}. For a prior much stronger than the data sets, we have $\widehat{\bm{\xi}}_{ABP} \approx \widehat{\bm{\xi}}_{SP}
\approx \widehat{\bm{\xi}}_{P}$, such that
\begin{equation}
\chi^2_{SP}(\widehat{\bm{x}}_{ABP}) - \chi^2_{SP,{\rm min}} \quad\to\quad
\chi^2_{S}(\widehat{\bm{\theta}}_{ABP},\widehat{\bm{\xi}}_{P}) - \chi^2_{S}(\widehat{\bm{\theta}}_{SP},\widehat{\bm{\xi}}_{P}) 
\qquad\text{(strong prior)}
\end{equation}
and the terms from the prior cancel. Hence, $\cPGpl$ reduces to the standard PG with $\bm{\xi}$ fixed to best fit from the prior alone. Therefore, it will follow a $\chi^2_p$ distribution. In contrast, for a prior much weaker than the $\bm{\xi}$ determination from the data sets, we have 
$\widehat{\bm{\xi}}_{ABP} \approx \widehat{\bm{\xi}}_{AB}$
and
$\widehat{\bm{\xi}}_{SP} \approx \widehat{\bm{\xi}}_{S}$, such that
\begin{equation}
\chi^2_{SP}(\widehat{\bm{x}}_{ABP}) - \chi^2_{SP,{\rm min}} \quad\to\quad
\chi^2_{S}(\widehat{\bm{x}}_{AB}) - \chi^2_{S}(\widehat{\bm{x}}_{S})
\qquad\text{(weak prior)} \,.
\end{equation}
In this case the prior terms cancel as distribution, because the prior terms will be dominated by the fluctuations of $\widehat{\bm{\xi}}_P$, and therefore $\Delta\chi^2_P(\widehat{\bm{x}}_{AB}) \sim \Delta\chi^2_P(\widehat{\bm{x}}_{S})$ and fully correlated. Hence, in this limit, $\cPGpl$ reduces to the PG for $A$ and $B$ without prior, which will be $\chi^2_{p+q}$ distributed. 

\section{Numerical results}
\label{sec:numres}

\subsection{Toy Monte-Carlo details}
\label{sec:MC}

The theoretical distributions derived in the previous sections can be validated against Monte Carlo (MC) data. As an example, let us consider a toy experimental setup inspired by a neutrino oscillation experiment. In particular, we will work in a two-family approximation, where the probability is fully determined by two parameters: $\Delta m^2 = m_2^2 - m_1^2$ ($m_i$ being the mass of each neutrino eigenstate) and a single mixing angle. In the flavor basis, the survival probability for a neutrino after propagating over a distance $L$ from the source, can be simply written as:
\begin{equation}
    P(U, \Delta m^2) = 1 - U \sin^2\left( \frac{\Delta m^2 L}{4 E_\nu} \right) \,,
\end{equation}
where $E_\nu$ stands for the neutrino energy, and $U \in [0,1]$ parametrizes the oscillation amplitude (which is a function of the mixing angle). For the purposes of this work it is useful to rewrite the probability in terms of the dimensionless parameter $r \equiv \Delta m^2 / \Delta m^2_0$ for some fixed reference value $\Delta m^2_0$, which yields
\begin{equation}
    P(U, r) = 1 - U \sin^2( r \Delta_0 ) \,,
\end{equation}
where $\Delta_0 \equiv \Delta m^2_0 L / (4E_\nu)$. Choosing $L$, $E_\nu$ and $\Delta m^ 2_0$ such that $\Delta_0 \sim \pi/2$ implies that $r\sim\mathcal{O}(1)$ in the region of optimal sensitivity to the oscillation probability, which is convenient. 

The distribution of signal events as a function of reconstructed neutrino energy can be computed in terms of the total neutrino flux $\phi$, the detection cross section ($\sigma_{\rm det }$, which depends on the neutrino energy) and the oscillation probability, as
\begin{equation}
\label{eq:dNdE}
    \frac{dN(U, r)}{dE_\nu} = \mathcal{N} \frac{d\phi}{dE_\nu} \sigma_{\rm det} P(U, r) \,,
\end{equation}
where $\mathcal{N}$ is an overall normalization constant (which depends, for example, on the number of targets for the neutrino interaction inside the detector and its geometrical acceptance) and for simplicity we have neglected energy-dependent detection efficiencies or smearing effects. The physics observable is the binned number of events as a function of neutrino energy, $N_i$, where $i$ refers to the bin index, which is obtained by integrating \cref{eq:dNdE} within the limits of each energy bin. In our simulations, we specifically consider a reactor neutrino flux (taken from Refs.~\cite{Murayama:2000iq, KamLAND:2002uet}) and an inverse beta-decay detection cross section (taken from Ref.~\cite{Vogel:1999zy}). The value of $\mathcal{N}$ is chosen such that we obtain hundreds of events (even after oscillations) for all bins, so the Gaussian approximation applies. The main parameters used for the configuration of our toy experiments $A$ and $B$, as well as the total event rates within the energy window used in the analysis, are summarized in Tab.~\ref{tab:toy-exps} for completeness. 

\begin{table}[t!]
    \centering
    \begin{tabular}{c|c c c c c}
    \hline
    Exp.  & $L$ (km) & Number of bins & $E_{\nu, \rm{min}}$  & $\Delta E_{\nu}$ & Unoscillated events \\ \hline 
      $A$   & 5.5 & 3 & 4.1 MeV & 0.2 MeV & 4,153 \\
      $B$   & 3.5 & 3 & 4.1 MeV & 0.2 MeV & 15,384 \\  \hline 
    \end{tabular}
    \mycaption{Main parameters used for the configuration of the toy experiments in our simulations. The energy bins are taken with equal width $\Delta E_\nu$, starting at $E_{\nu, \rm{min}}$. The total number of events reported has been obtained without oscillations (that is, for $U=0$). No backgrounds or systematic uncertainties are considered in our analysis. }
    \label{tab:toy-exps}
\end{table}

Given the values of $L$ and $E_\nu$ in \cref{tab:toy-exps}, oscillations are maximized around $\Delta m^2 \sim 2.5\times 10^{-3}~\mathrm{eV}^2$. Thus, $\Delta m^2_0 = 10^{-3}~\mathrm{eV}^2$ is a natural choice, and unless otherwise noted we will take $\overline r=2.5$ as our true value. Furthermore, in order to ensure that the value of $U$ always remains away from the physical boundaries at $U=0, 1$ we set its true value at $\overline U = 0.587$. 

Once we have defined our physical observables, we may consider two toy experiments $S\equiv A,B$, each of aiming to determine the oscillation parameters $\theta = \left\{ r, U \right\}$ from their observed data $O^S_i$. Within the Gaussian approximation, we may define a binned $\chi^2$ function for each of them as 
\begin{equation}
\chi^2_{S}(\theta, \xi) = \sum_i \left(\frac{O^S_i - N^S_i(\theta, \xi)}{\sigma^S_i}\right)^2
    \label{eq:chi2-gauss}
\end{equation}
where $\sigma_i^S = \sqrt{O^S_i}$ is the statistical error in each bin. The covariance matrices for experiments $A$ and $B$ (defined in \cref{tab:toy-exps}), computed at the true input values for $r$ and $U$, read:
\begin{equation}
    V_A \simeq \begin{pmatrix}
        3.5 & -5.1    \\
       -5.1 &  7.8	
    \end{pmatrix} \times 10^{-3}; \quad
    V_B \simeq \begin{pmatrix}
       3.7  & 5.7 \\
       5.7  & 9.2
    \end{pmatrix} \times 10^{-3} \, . 
    \label{eq:VA-VB}
\end{equation}

The prior constraint is implemented in our simulation through the addition of quadratic pull terms, as in \cref{eq:chisqP}:
\begin{eqnarray}\label{eq:prior-MC}
    \Delta\chi^2_U = \frac{(U - \widehat{U})^2}{s_U} \,, \qquad 
    \Delta\chi^2_r = \frac{(r - \widehat{r})^2}{s_r} \,.
\end{eqnarray}

In order to derive the expected distribution for our toy experiments, we need to simulate a large sample of pseudo-data in presence of statistical fluctuations: these will be used as $O_i$ in \cref{eq:chi2-gauss}. In doing so, we first pick a set of ``true'' oscillation parameters and nuisance parameters $\{\bar\theta, \bar\xi\} $ and compute the predicted event rates $N_i(\bar\theta, \bar\xi)$. Then, random fluctuations are generated around the predicted number of events in each bin
assuming a Gaussian distribution with width $\sigma_i = \sqrt{N_i}$. In full analogy, a similar procedure should be followed for the priors since these are also obtained from data (and are therefore subject to statistical fluctuations). After picking a set of true values for each prior ($\bar \xi$), for each pseudo-experiment a random number is generated around $\bar\xi$ using a Gaussian distribution with a width according to the prior experimental uncertainty. The resulting value, $\widehat{\xi}_P$, will be used as central value for the pull term in \cref{eq:prior-MC}. As outlined in previous sections, the resulting distributions will depend on the number of nuisance parameters included in the fit. Therefore, for our simulations we have considered the following two scenarios: 
\begin{itemize}
    \item We first assume that there is a prior measurement only on $U$, which now acts as a nuisance parameter in our fit. Hence, $p=1,\, q=1$ in this case, with $\theta = r$ and $\xi = U$, we only impose $\Delta\chi^2_U$ from \cref{eq:prior-MC} and have $s_P=s_U$. 
    \item Second, we consider a case with two prior measurements imposing both terms from \cref{eq:prior-MC}, that is, $p=0,\, q=2$, with $\bm{\xi} = (r,U)$. In this case $S_P$ is a $2\times 2$ diagonal matrix with entries $s_r$ and $s_U$, \textit{i.e.}, we assume the prior measurements to be uncorrelated. 
\end{itemize}

\subsection{Results for the null hypothesis}
\label{sec:H0}

\begin{figure}[t]
  \centering
  \includegraphics[width=0.9\textwidth]{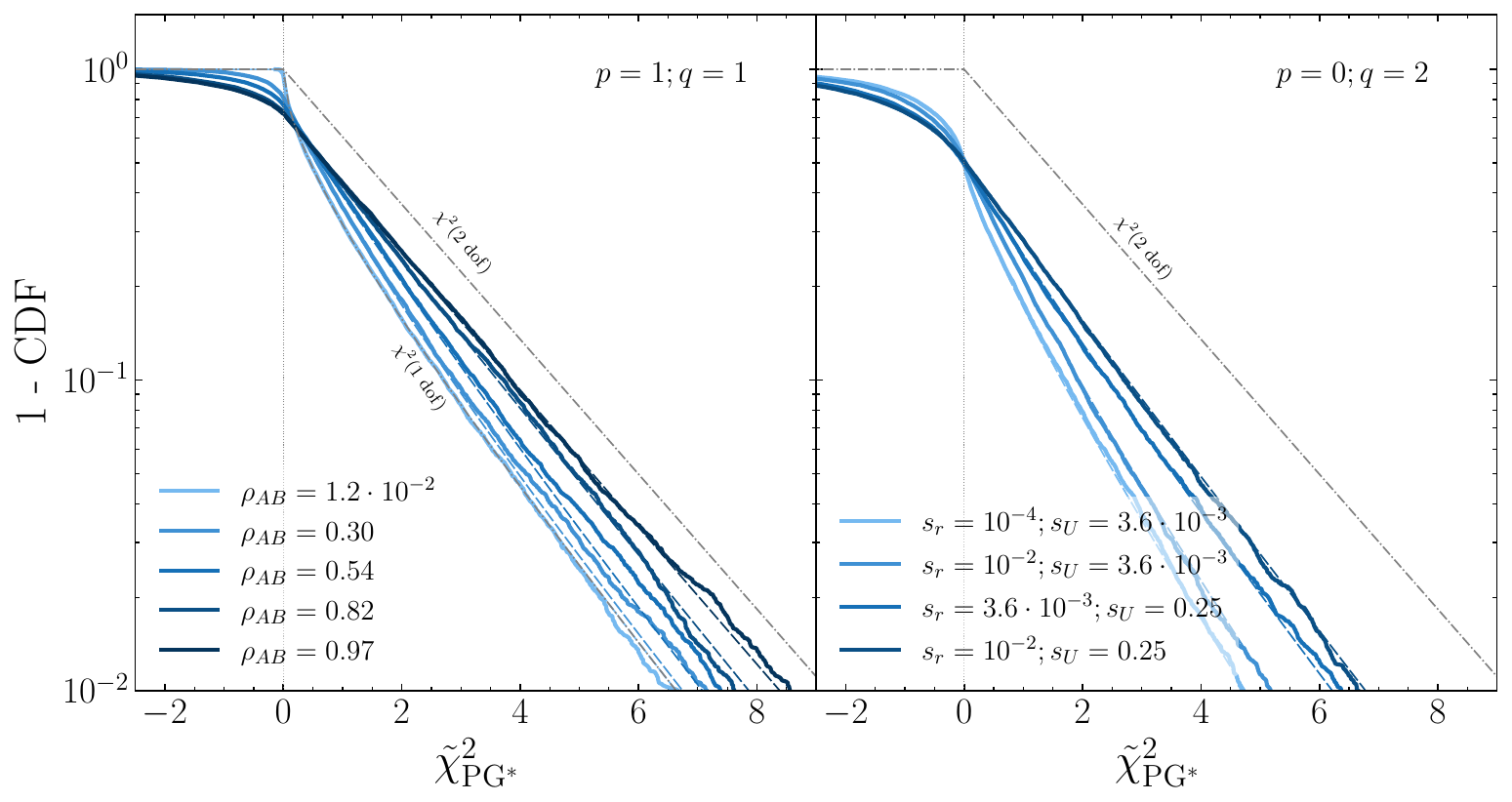}
  \mycaption{Complementary cumulative distribution function of the \PGst\ test statistic. Thick solid lines show the numerical results obtained for $10^4$ pseudo-experiments, while thin dashed lines show the theoretical distributions obtained in \cref{sec:PGst}. Dash-dotted gray lines show the distributions obtained for a $\chi^2$ with one and two degrees of freedom. The left panel corresponds to the case $p=1,\, q=1$, and we show curves for different values of the correlation parameter $\rho_{AB}$ defined in \cref{eq:rho}. The right panel corresponds to the case $p=0,\, q=2$, and we give the values for the prior variances $s_U$ and $s_r$ in the legend, see \cref{eq:prior-MC}.}
  \label{fig:cdf}
\end{figure}

We present now the the numerical results of our simulations, comparing them to the theoretical distributions derived for the test statistics in \cref{sec:PGst,sec:PGpl}.
\Cref{fig:cdf} shows our results for the \PGst\ test statistic, where we compare the complementary CDF calculated from simulation (thick solid lines) to the theoretical distributions in \cref{eq:cdf} (thin dashed lines). The left panel shows our results for $p=1,\, q=1$ (that is, when only a prior on $U$ is imposed) while the right panel shows the corresponding result for $p=0,\, q=2$ (when priors are imposed on both $U$ and $r$). In both panels, the different lines illustrate how the results change as a function of the priors, between the strong prior limit ($s_P \to 0$, light blue lines) and weak prior limit ($s_P \to\infty$, dark blue lines). 

In all cases the agreement between the numerical results and the theoretical distributions is excellent, with mild differences that can be attributed to the limited statistics of the simulated sample of pseudo-data. We also see that, for all lines, the random variable corresponding to the \PGst\ admits negative values (as expected). As can be seen from the figure, for $\rho_{AB} \lesssim 0.2$ the distribution remains reasonably close to a $\chi^2_1$, whereas for $\rho_{AB} \gtrsim 0.2$ relevant deviations appear. \Cref{fig:cdf} also shows how in the weak prior limit the obtained distributions do not converge towards a normal $\chi^2$; instead, they remain a generalized $\chi^2$ distribution, as anticipated in \cref{sec:PGst}.

\begin{figure}[t]
  \centering
  \includegraphics[width=0.9\textwidth]{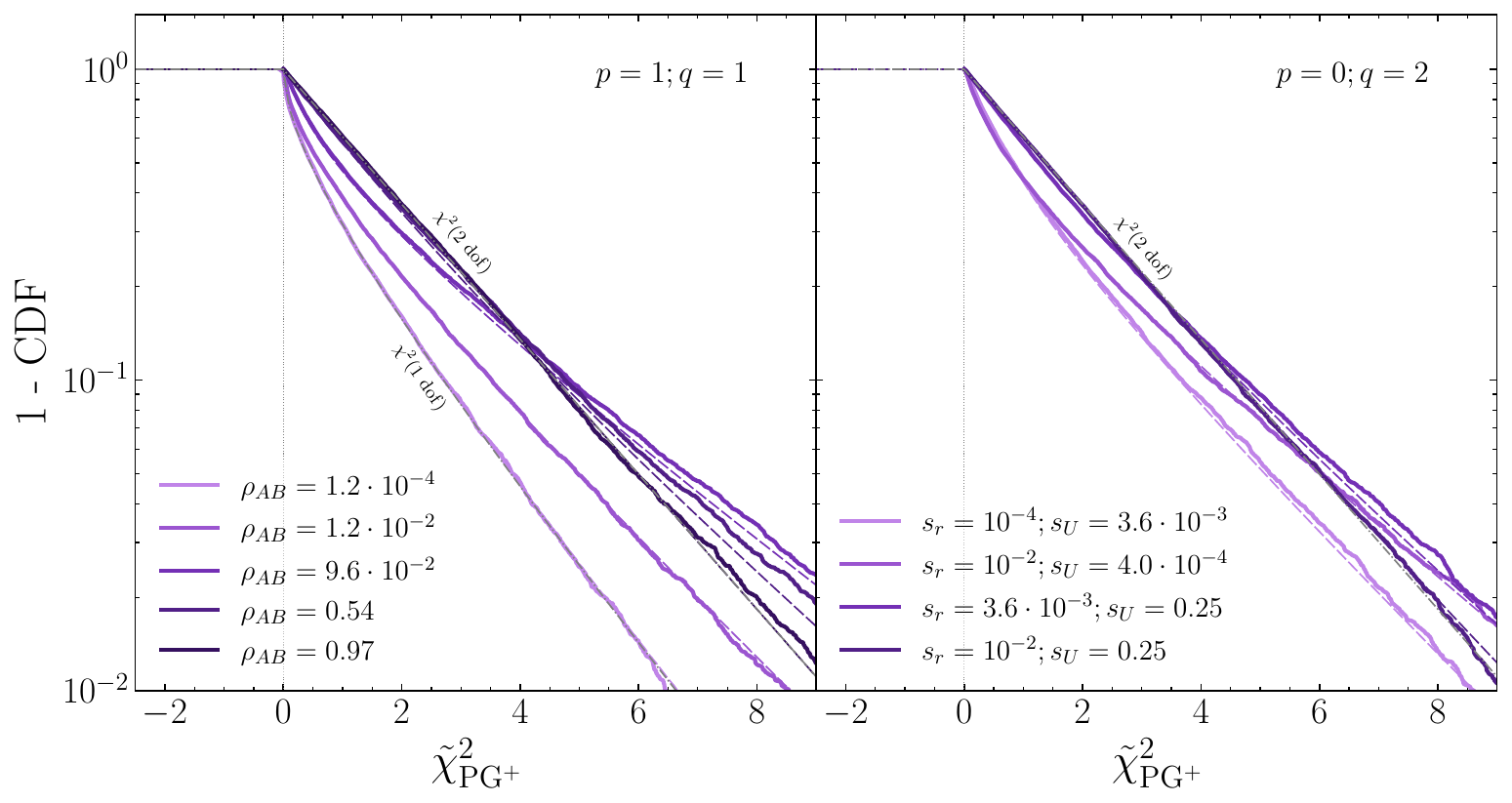}
  \mycaption{Complementary cumulative distribution function of the \PGpl\ test statistic.
 Thick solid lines show the numerical results obtained for $10^4$ pseudo-experiments, while thin dashed lines show the theoretical distributions obtained in \cref{sec:PGpl}. Dash-dotted gray lines show the distributions obtained for a $\chi^2$ with one and two degrees of freedom. The left panel corresponds to the case $p=1,\, q=1$, and we show curves for different values of the correlation parameter $\rho_{AB}$ defined in \cref{eq:rho}. The right panel corresponds to the case $p=0,\, q=2$, and we give the values for the prior variances $s_r$ and $s_U$ in the legend, see \cref{eq:prior-MC}.
  }
  \label{fig:cdf-PGpl}
\end{figure}

Our results for the \PGpl\ test statistic are shown in \Cref{fig:cdf-PGpl}, where the complementary CDF calculated from our simulation (thick solid lines) is compared to the theoretical distributions obtained in \cref{sec:PGpl} (thin dashed lines), for $p=1,\, q=1$ (left panel) and $p=0,\, q=2$ (right panel). Again in this case, the agreement between our numerical results and the theoretical distributions is excellent. Darker/lighter lines show how the results change as we approach the weak/strong prior limits, respectively. As discussed in \cref{sec:PGpl}, for $p=1,\, q=1$ in the strong prior limit ($s_P \to 0$) the distribution matches a $\chi^2$ with one degree of freedom. 
However, the obtained distributions start to show large deviations from the $\chi^2_1$ limit even for small $\rho_{AB} \sim \mathcal{O}(0.01)$. This contrasts with the behaviour of the \PGst\ test, where a $\chi^2_1$ remains a relatively good approximation even for $\rho_{AB}$ as large as 0.2. \Cref{fig:cdf-PGpl} also shows that for intermediate values of $s_P$ the tail of the distribution can go above that of a $\chi^2_2$; however, as the strength of the prior approaches the weak limit, the distribution converges smoothly to a $\chi^2_2$, as anticipated in \cref{sec:PGpl}.

Note that for the case $p=0$ shown in the right panels of \cref{fig:cdf,fig:cdf-PGpl}, the limit of infinitely strong prior would formally lead to a $\chi^2$ distribution with zero degrees-of-freedom, meaning no randomness. In this case, the prior-constrained PG tests would measure just the consistency of the prior with itself, which by construction is perfect. 

Let us mention that our toy MC is on purpose constructed such that we expect the Gaussian approximation to hold (high-statistics data, true parameter values far away from boundaries, and only one minimum in the $\chi^2$).  A detailed study of a real-life setup using actual neutrino oscillation data, where some of these properties do not hold, will be presented in Ref.~\cite{inpreparation}.  

\paragraph{Comparison between the two test statistics.}

\begin{figure}[t]
  \centering
  \includegraphics[width=0.85\textwidth]{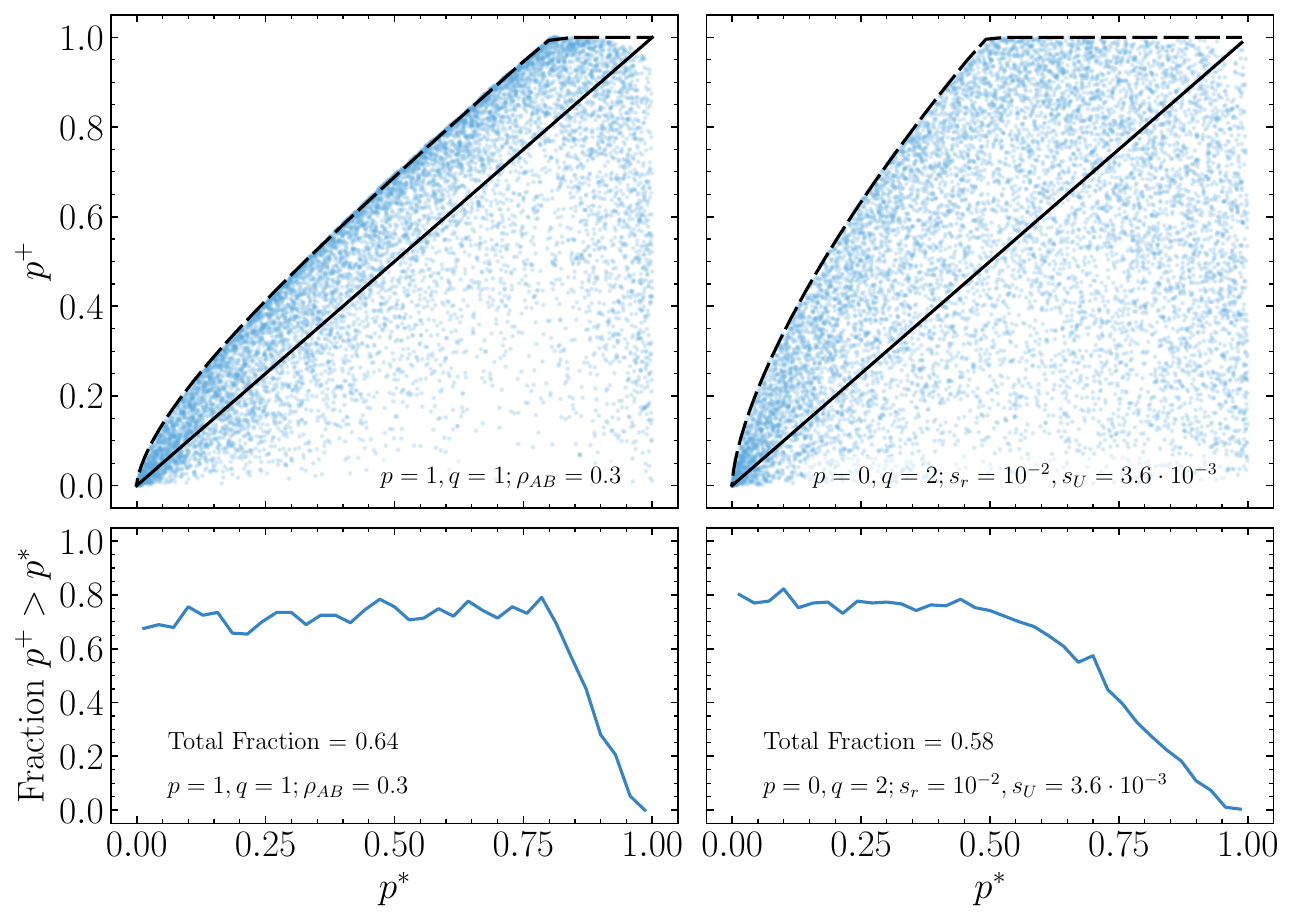}
  \mycaption{Comparison between the $p$-values obtained for the \PGst\ ($p^*$) and \PGpl\ ($p^+$) test statistics, for two examples with $p=1,\, q=1$ (left panels) and $p=0,\, q=2$ (right panels). Upper panels: scatter plot of the $p$-values obtained for the two tests statistics considered, for $10^4$ experimental realizations (red points). The diagonal black line corresponds to $p^+=p^*$, while the dashed black line shows the curve corresponding to \cref{eq:scatter-bound}, see text for details. Lower panels: fraction of points for which $p^+ > p^*$ as a function of $p^*$ for the same simulations.  
  }
  \label{fig:scatter}
\end{figure}

Since the two test statistics considered here differ on the way the prior is introduced (with a larger relative weight for \PGpl\ than for \PGst), a relevant question to ask is whether one of the two test statistics is more sensitive to tension in the fit than the other. In \cref{fig:scatter} we show the distribution of $p$-values for the two tests from our MC simulation, where 
we use the shorthand notation $p^*$ ($p^+$) for the $p$-value of \PGst\ (\PGpl). Again, the left panel corresponds to $p=1,\, q=1$ while the right panel shows the results for $p=0,\, q=2$ as indicated by the labels. 

The upper panels in \cref{fig:scatter} show the $p$-values obtained for the two tests statistics considered, for $10^4$ toy MC realizations. As shown in the upper panels in \cref{fig:scatter}, the points tend to cluster in the region above the diagonal, indicating that under the null-hypothesis it is likely to obtain a larger $p$-value for \PGpl\ than for \PGst. Analytically, this behaviour can be understood since statistical fluctuations can only make $\cPGpl \geq \cPGst$ by construction,
see \cref{eq:PGpl2}, which translates into a strict upper bound on $p^+$. Specifically, using the shorthand notation $z^* = \cPGst $ and $z^+ = \cPGpl$, it is easy to see that
\begin{equation}
     z^+ \geq z^* 
     \quad\Rightarrow \quad
     p^+ = \bar F^+(z^+) \leq \bar F^+(z^*)  \, ,
\end{equation}
where $\bar F^{+/*}(z) = 1 - F^{+/*}(z)$ refers to the complementary CDF for the \PGpl/\PGst\ test statistic, and we have $p^{+/*}=\bar F^{+/*}(z)$ by definition. Therefore, the upper bound on $p^+$ can be obtained by the curve 
\begin{equation}
    \bm{\gamma}(z) = \left[\bar F^*(z), \bar F^+(z)\right] \, ,
    \label{eq:scatter-bound}
\end{equation}
parametrized by $z$. The curve from \cref{eq:scatter-bound} is shown by the dashed black lines in the upper two panels in \cref{fig:scatter}. As can be seen, the agreement with our numerical results is again excellent. 
We have verified that in the strong prior limit (for $p>0$), the distribution of $p^*$ and $p^+$ values concentrate to points close to the diagonal $p^* = p^+$, as in that limit both the \PGst\ and \PGpl\ tests become identical and converge towards the standard PG.

In the same figure, the lower panels show the fraction of points where $p^+ > p^*$ as a function of $p^*$, for the same data sample. As can be seen, this fraction approaches a constant value around $\sim 0.7-0.8$ in the region $p^* \lesssim \mathcal{O}(0.75)$, while for larger values of $p^*$ it drops to zero. We note that the fraction remains at $\sim 0.7-0.8$ for small values of $p^*$ regardless of the prior strength; however the exact value where the fraction drops depends on the number of priors considered and the values assumed, as illustrated by the two examples shown in \cref{fig:scatter}.  The value of $p^*$ where the fraction starts to drop is approximately set by the $p$-value obtained for $\cPGst=0$, where the dashed curve from \cref{eq:scatter-bound} reaches $p^+ = 1$. In fact, as $\cPGst$ decreases towards negative values, the upper bound on $p^+$ does not change (since it had already saturated at 1) but $p^*$ keeps increasing, and therefore the fraction of points where $p^+ > p^*$ drops. 

Hence, for not too large $p$-values, we find that under the null-hypothesis it is likely to find a larger $p$-value for \PGpl\ than for \PGst. As the prior has more weight in the \PGpl\ test statistic, this indicates that it is somewhat more likely that fluctuations contribute to the $\chi^2$-terms involving the data sets $A$ and $B$ rather than to the prior term $\Delta\chi^2_P$. Comparing the values of $p^*$ and $p^+$ for a given realisation of the data may be used as an additional diagnostic tool to identify the origin of a tension. 

\subsection{Results for alternative hypotheses}
\label{sec:H1}

Until now we have focused on the computation of the expected distributions under the null hypothesis $H_0$. It assumes that the model is true, \textit{i.e.}, that the two experiments as well as the prior measurement are random realisations of the data for the same true values of the underlying model parameters.
However, it is also worth considering the distribution of the test statistics under an alternative hypothesis $H_1$. 

The theoretical distribution for an alternative hypothesis can be derived following the same procedure as in \cref{sec:PGst,sec:PGpl} bearing in mind that, in this case, the variables will no longer be normally distributed around zero but around some non-vanishing value instead (which eventually depends on the alternative hypothesis chosen). For example, if we consider a hypothesis where an experiment $S$ and the prior $P$ observe data that are generated with different true values of the nuisance parameter, $\overline\xi_S$ and $\overline \xi_P$, then $\chi^2_{\mathrm{PG}, SP} = d_{SP}^2$ will follow a non-central chi-squared distribution with one d.o.f.\ and non-centrality parameter $(\overline\xi_S - \overline \xi_P)^2/(s_S + s_P)$. Likewise, it can be shown that $\chi^2_{\mathrm{PG}, ABP}$ will follow a non-central distribution with 3 d.o.f., where the statistically independent variables can be obtained through the same change of basis as in \cref{sec:PGst,sec:PGpl}. Therefore, both \PGst\ and \PGpl\ can be written as a sum of non-central chi-squared distributions which, by definition, follows a generalized $\chi^2$ distribution~\cite{wiki, imhof:1961}. Our accompanying code~\cite{ref-code} provides auxiliary functions to compute the non-centrality parameters (as well as the resulting distribution) for any choice of true values for the underlying model parameters. 

As outlined above, the choice of alternative hypothesis is not unique. While a systematic study of this is beyond the scope of this work, we have identified two general cases that are physically motivated:
\begin{description}
    \item[Case I:] both data sets $A$ and $B$ are based on the same set of true model parameters while the prior measurement is generated from different true values. In this case the tension in the combined fit arises from the mismatch between the prior and the data sets.  
    \item[Case II:] the true value of the prior is in agreement with the one of $U$ used to simulate the data, but different true values of $r$ are used for experiments $A$ and $B$. In this case the prior is consistent with both experiments separately; however, when analysed in combination a tension emerges in the fit.
\end{description}
We are mostly interested in a scenario where there is significant tension between the different data sets and/or the prior. This implies that the best-fit points for the joint fit will be significantly displaced from the minimum obtained for the Asimov data set for each experiment, which could eventually make our results sensitive to non-linearities entering the $\chi^2$. Thus, in order to guarantee that our simulations still fall within the linear regime also for the alternative hypotheses, the results obtained in the remainder of this section have been obtained multiplying by 30 the statistics of the experiments with respect to our default setups defined in \cref{sec:MC}.

\begin{figure}[t]
  \centering
  \includegraphics[width=0.49\textwidth]{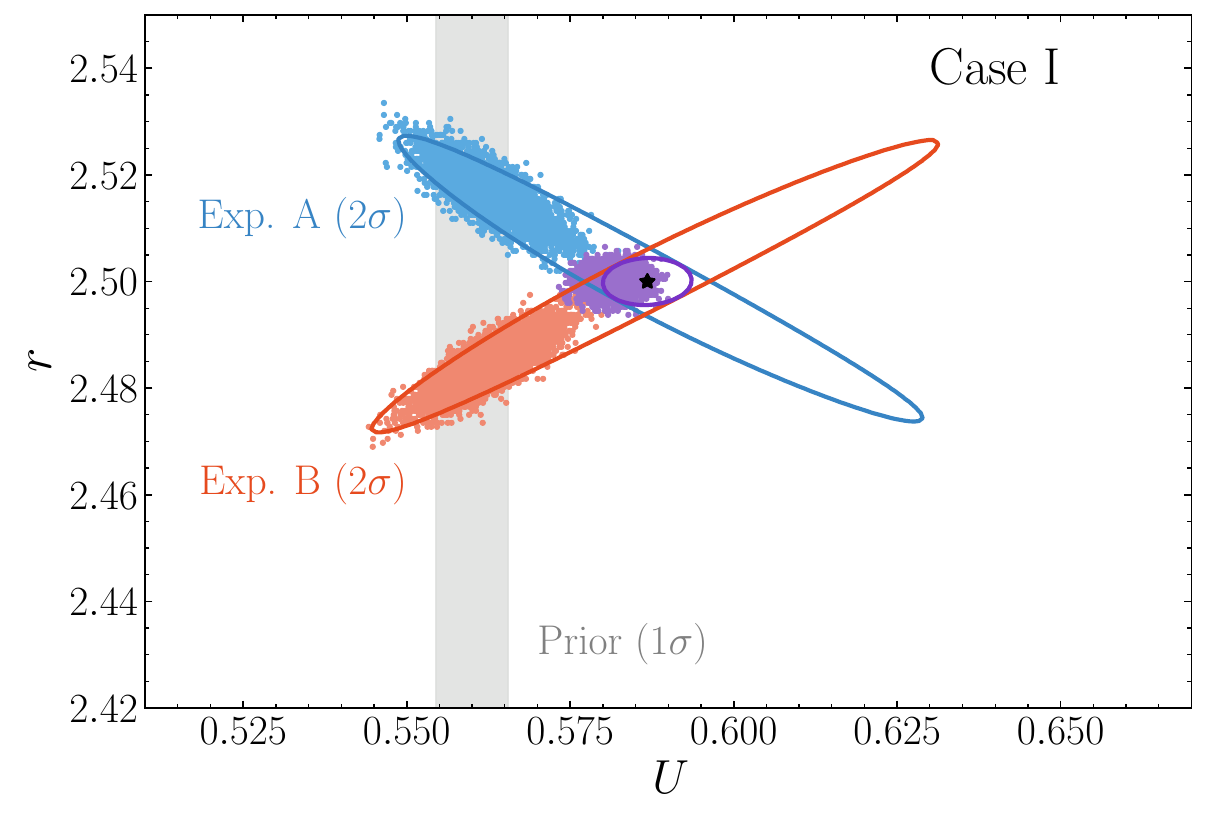}
  \includegraphics[width=0.49\textwidth]{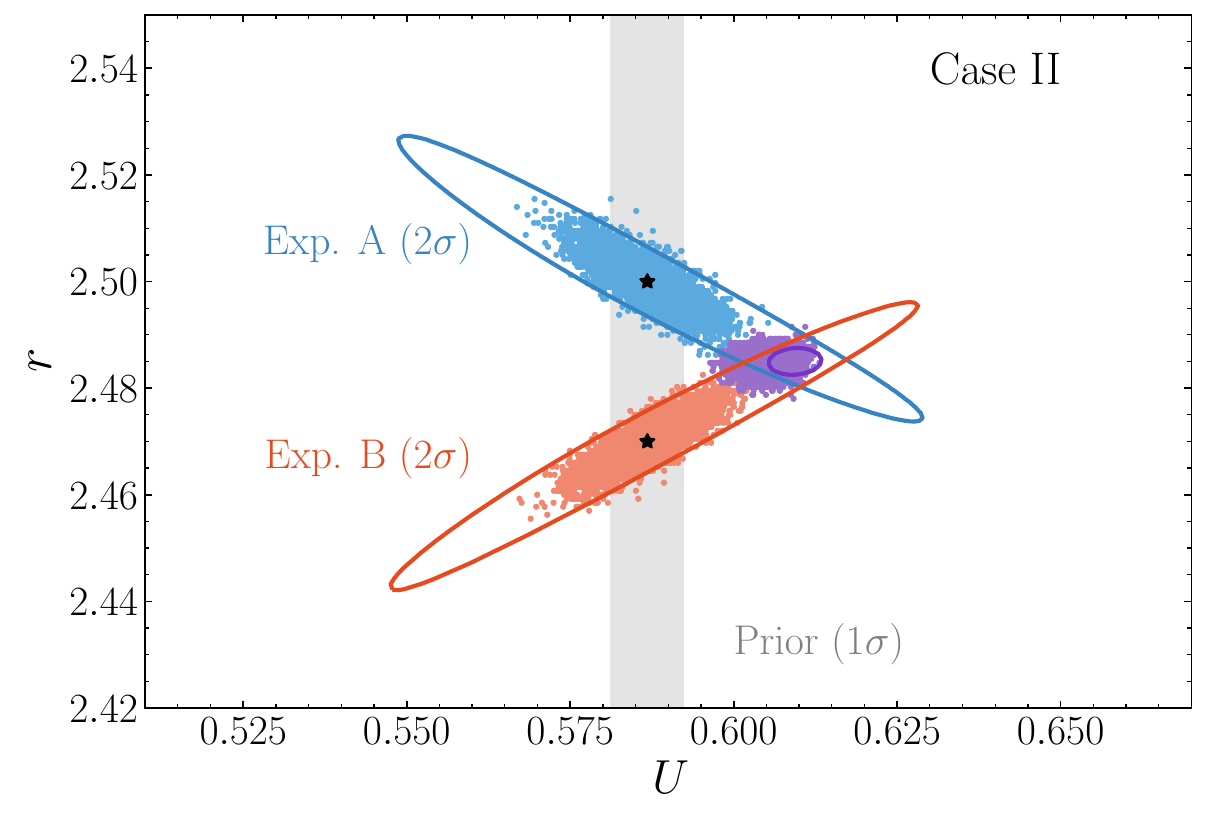}
  \mycaption{Example showing the sensitivity regions at 2$\sigma$ CL (for 2~d.o.f.) for the two experiments under consideration. The star indicates the true values of the parameters used to simulate each experiment, and the solid contours show the sensitivity regions obtained with the Asimov data set for each experiment separately (blue/red lines) and in combination (purple line). The dots indicate the location of the global minima found when pseudo-data for the two experiments are fitted separately (blue and red dots for experiments A and B, respectively) or jointly (purple dots), in all cases imposing the same prior on $U$. Left panel (case I): the best-fit of the prior measurement is set to a different value of $U$, as indicated by the gray band. Right panel (case II): the best-fit of the prior measurement is in agreement with the value of $U$ used to simulate the data, but different true values of $r$ are used for experiments $A$ and $B$. }
  \label{fig:asimov}
\end{figure}

\begin{figure}[t]
  \centering
  \includegraphics[width=0.95\textwidth]{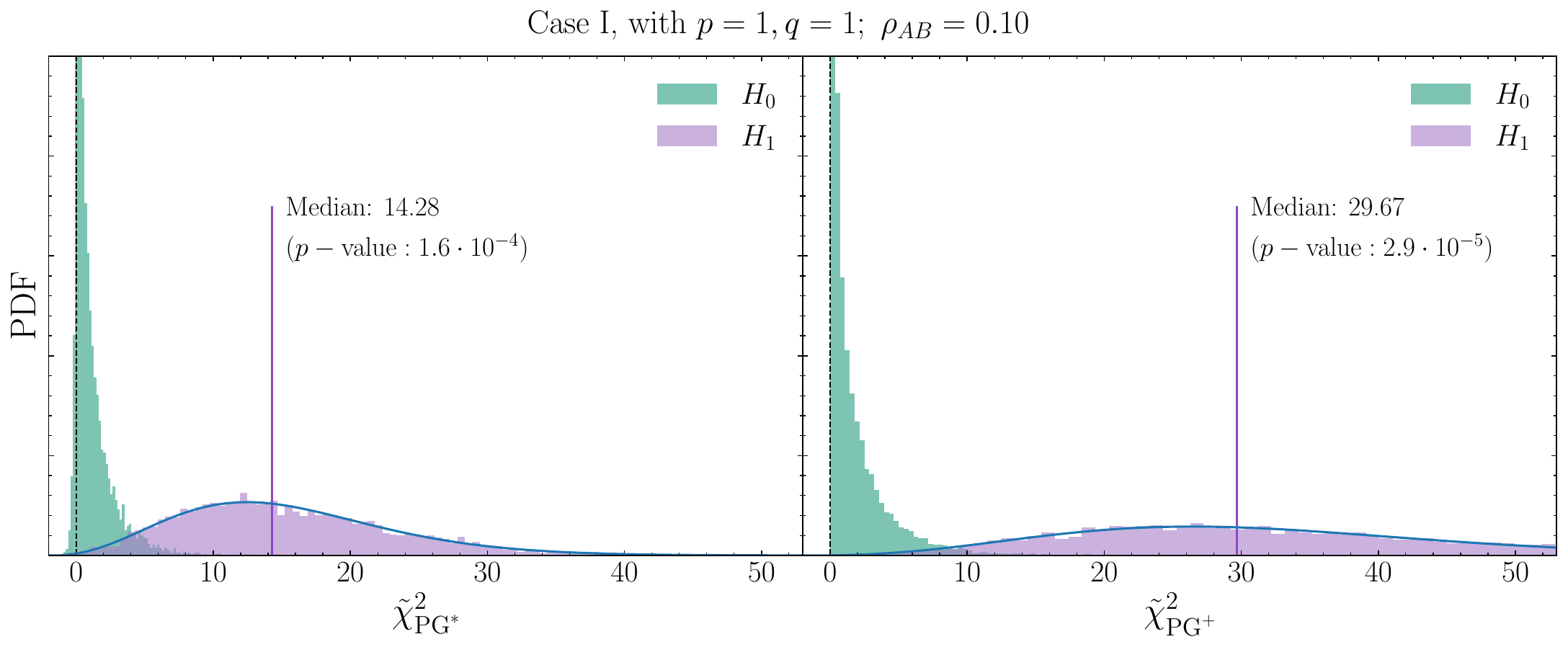}
  \mycaption{Example showing the distributions of pseudo-experiments for the alternative hypothesis ($H_1$, purple histograms), compared to the null hypothesis ($H_0$, teal histograms). The solid blue curve on top of the purple histograms shows the theoretical distributions, in excellent agreement with our simulation. The vertical purple lines indicate the sensitivity of an average experiment, see text for details. This example corresponds to case I; the results for case II are qualitatively very similar.  }
  \label{fig:median-sens}
\end{figure}

\Cref{fig:asimov} shows the interplay of the different data sets in the combined fit for two specific examples satisfying the criteria outlined above. In both panels, the best-fit points for each experiment separately (blue and red dots, for $A$ and $B$, respectively) do not show any tension: they are found to be in good agreement with the sensitivity obtained using the Asimov data as well as with the prior. Conversely, a tension is observed for the joint fit (purple dots) in both panels. The distributions of pseudo-experiments for the null and alternative hypotheses are shown in \cref{fig:median-sens} taking case I as an example, for the two test statistics under consideration. We find that the theoretical distributions obtained for the alternative hypothesis (solid blue lines) are in excellent agreement with our numerical results. Note that the results for case II are qualitatively very similar and are therefore not shown here.

\paragraph{Median sensitivity.} The tension observed in \cref{fig:asimov} can be further quantified by the $p$-value corresponding to the median of the distribution of pseudo-experiments for the alternative hypothesis. This defines the sensitivity of an average experiment or \emph{median sensitivity}. In \cref{fig:median-sens}, in each panel, the vertical purple lines indicates the $p$-value corresponding to the median sensitivity. For our alternative hypotheses we find: 
\begin{align}
    \left.
    \begin{aligned}
        p^*|_{\rm median}  &= 1.6\cdot 10^{-4} ~(3.6\sigma) \\
        p^+|_{\rm median} &= 2.9\cdot 10^{-5} ~(4.0\sigma)
    \end{aligned}
    \right\} \quad &\text{for case I}  \label{eq:median-I}
    \qquad \\
    \left.
    \begin{aligned}
        p^*|_{\rm median} &= 2.6\cdot 10^{-5} ~(4.1\sigma)\\
        p^+|_{\rm median} &= 3.8\cdot 10^{-5} ~(4.0\sigma)
    \end{aligned}
    \right\} \quad &\text{for case II}. \label{eq:median-II}
\end{align}
where the reported number of standard deviations corresponds to a one-sided Gaussian. We see that for case I we obtain a somewhat larger tension for \PGpl\ than for \PGst. This is expected, since in this case the tension emerges from the prior itself, which has a larger relative weight for \PGpl\ than for \PGst. Conversely, for case II the difference in $p$-values is reduced, with a slightly lower tension for \PGpl\ than for \PGst. This is again expected since the tension comes from the data instead of the prior and, moreover, the relative weight of the data is lower for \PGpl.

Overall, the example provided above illustrates how a comparison between the $p$-values obtained for \PGst\ and \PGpl\ may aid in determining the source of a possible tension in the data, as previously mentioned in \cref{sec:H0}. 

\paragraph{Power of the test.}
Until now we have focused on the computation of the type-I error rate associated to a given test statistic, that is, the probability to reject the null hypothesis $H_0$ even though it is true (\textit{i.e.,} the $p$-value). However, the type-II error rate $\beta$ is also relevant: it is defined as the probability of not rejecting $H_0$ although it is false. In order to compute it, we first need to pick a (true) alternative hypothesis and obtain its corresponding distribution for the test statistic. Next, we choose a $p$-value and determine the associated value of the test statistic under $H_0$. The corresponding type-II error can be obtained as the fraction of experiments within $H_1$ which would yield a lower value. Then, $1-\beta$ is denoted as the \emph{power} of the test. 
 
\begin{figure}
    \centering
    \includegraphics[width=0.95\linewidth]{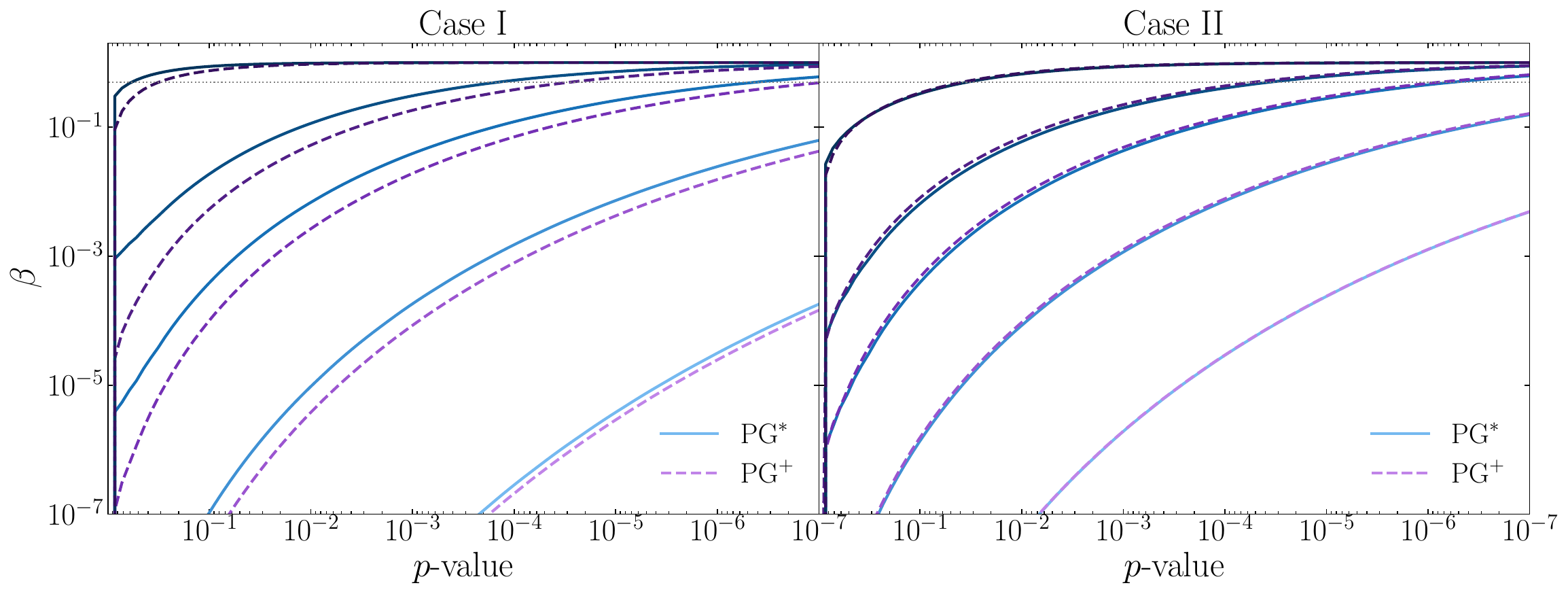}
    \mycaption{Type-II error rate of the test as a function of the $p$-value, for the two test statistics under consideration. Left and right panels show the results for alternative hypotheses matching cases I and II, respectively. In each panel, each pair of lines has been obtained for a different prior, between $\rho_{AB} = 3.5\cdot 10^{-3}$ (lightest) to $\rho_{AB} = 0.6$ (darkest). }
    \label{fig:power}
\end{figure}
\Cref{fig:power} shows the type-II error rate $\beta$ of the test statistics as a function of the $p$-value for the two cases considered here. By definition, the $p$-value corresponding to $\beta=0.5$ (indicated by a horizontal dotted line) yields the median sensitivity in \cref{eq:median-I,eq:median-II}. However, the relative difference in the tension reported by the two test statistics is also present for other values of $\beta$: while for case I (left panel) the two curves are more separated with \PGpl\ having slightly more power (\textit{i.e.,} lower $\beta$), for case II (right panel) the two test statistics behave in a very similar manner.\footnote{In case~I the tension arises from the prior. Since the prior has more weight in the \PGpl\ statistics, it has a somewhat better power than \PGst\ in this case. In case~II the tension emerges from the data, and \PGst\ gives slightly better power than (though very similar to) \PGpl.} 
In the same figure we also show how the curves change as a function of $\rho_{AB}$. We find that the relative comparison between the two sets of curves generally holds as $\rho_{AB}$ is varied, except when the weak prior limit is approached. This is expected since, in the weak prior limit, the tension from the prior disappears and therefore $\beta \to 1$ regardless of the $p$-value (that is, the discrimination between the two hypothesis is no longer possible for the particular examples chosen here). We also see that in the strong prior limit the two lines become very similar even for case I, since in this limit both tests converge to the standard PG.

\section{Summary}\label{sec:summary}

In this work we have proposed an extension of the Parameter Goodness-of-fit (PG) test, which can be used when some of the parameters of the model are constrained by a prior common to the data sets. In such a case, the data and the prior are no longer statistically independent, so the original PG test cannot be applied. We have proposed two modified versions of the PG test statistic, denoted by \PGst\ and \PGpl. Both can be used to test the compatibility of data sets subject to a common prior, so we call the corresponding test \emph{prior-constrained parameter goodness-of-fit}.

The two test statistics differ by a term involving the prior $\chi^2$. The \PGpl\ is constructed to be strictly non-negative thanks to the additional prior term, whereas \PGst\ can assume also negative values and gives somewhat less weight to the prior. Both test statistics converge towards the standard PG test in the case of ``strong prior'', \textit{i.e.}, when the constraint on the parameters from the prior is much stronger than the one from the data sets, with a number of degrees of freedom that is determined by the number of parameters unconstrained by the prior.  

In the Gaussian approximation the \PGst\ and \PGpl\ test statistics are distributed according to a so-called generalised $\chi^2$ distribution, which can be easily calculated via its characteristic function. The distribution is determined just by the parameter covariance matrices of the data sets and the prior. Once the distribution is known, $p$-values can be calculated in a straightforward manner. While the distribution of the test statistics depend on the prior strength, the expectation value of the \PGst\ test statistic is given by the numbers of parameters unconstrained by the prior, $p$, independently of the prior strength, see \cref{eq:expPGst}. Hence a value of the test statistic much larger than $p$ indicates a low $p$-value. For the \PGpl\ test statistic the expectation value depends on the prior strength and varies between $p$ for a strong prior till $p+q$ for a weak prior, where $p\, (q)$ is the number of parameters unconstrained (constrained) by the prior. We provide an accompanying Python code \cite{ref-code} which allows to calculate the corresponding distribution functions and $p$-values for any number of model and nuisance parameters, given a set of covariance matrices. 

Our results have been validated through Monte Carlo simulation of two toy experiments inspired by neutrino oscillation facilities. The simulated setup is close to the Gaussian limit (large statistics, no multiple minima, and away from physical boundaries). The simulations show excellent agreement with the analytic expressions for the distributions of the \PGst\ and \PGpl\ test statistics.  We find that, under the null hypothesis and for not too large $p$-values, it is more likely to obtain a larger $p$-value for \PGpl\ than for \PGst. This indicates that it is more likely that statistical fluctuations contribute to the $\chi^2$ of individual experiments than to the prior term. A comparison between both $p$-values may be used as an additional tool to diagnose the origin of a tension. Finally, we have also computed the power of the test for an alternative hypothesis, for two physically motivated examples, illustrating that our test statistics are able to detect tensions between data sets in the presence of a common prior. We find that both test statistics have comparable power, with \PGpl\ performing somewhat better, especially in cases where the tension is mostly driven by the prior. 

In summary, we believe that the test statistics proposed here offer a useful generalisation of the PG test for analysing the statistical compatibility of data sets, and extends its range of applicability.

\subsection*{Acknowledgements}

We thank Yufeng Li for asking a relevant question about the PG test, which triggered this work. We acknowledge the help of ChatGPT and Claude for developing code as well as for some of the mathematical derivations. This project is funded by the European Union's Horizon Europe research and innovation programme (Marie Sk{\l}odowska-Curie Staff Exchange grant agreement 101086085-ASYMMETRY) and by ERDF ``A way of making Europe''.  It also receives support from the R\&D\&I project CEX2025-001574-S ``Centro de Excelencia Severo Ochoa'' (the research presented in this publication falls within the research line ``Particle Physics in the Standard Model and Beyond (BSM)'') and from grants PID2022-\allowbreak 142545NB-\allowbreak C21 and PID2025-\allowbreak 171019NB-\allowbreak C21, all funded by MICIU/AEI/\allowbreak 10.13039/\allowbreak
501100011033.

\appendix

\section{Prior-constrained PG for more than two data sets}
\label{appendix}

Similar as the original PG test, also the prior-constrained PG can be extended in a straight-forward way to the case of more than two data sets with a common prior. We consider $N$ data sets, with $\chi^2_{SP}(\bm{\theta},\bm{\xi}) = \chi^2_S(\bm{\theta},\bm{\xi}) + \Delta\chi^2_P(\bm{\xi})$ for $S=1,\ldots,N$, and define
\begin{align}
  \cPGst &=
\min_{\bm{\theta},\bm{\xi}}
 \left[ \sum_{S=1}^N \chi_S^2(\bm{\theta},\bm{\xi}) + 
 \Delta\chi_P^2(\bm{\xi})\right]
 - \sum_{S=1}^N \chi^2_{SP,{\rm min}} \,,\label{eq:PGstmulti}
 \\
 \cPGpl &=
 \sum_{S=1}^N \Big[ \chi_{SP}^2(\widehat{\bm{x}}_{\rm all}) 
 - \chi^2_{SP,{\rm min}} \Big]
 = \cPGst + \big( N-1 \big)
 \Delta\chi^2_P(\widehat{\bm{\xi}}_{\rm all})
\end{align}
where $\widehat{\bm{x}}_{\rm all} = (\widehat{\bm{\theta}}_{\rm all}, \widehat{\bm{\xi}}_{\rm all})$ denotes the best fit parameters from the combination of all $N$ data sets and the prior $P$. Notice that, as already seen in \cref{eq:PGpl2}, $\cPGst$ and $\cPGpl$ differ just by a prior term.

We outline here the main idea for \PGst\ for the simplest case of $p=q=1$, and leave the generalization to the multi-parameter case as well as for \PGpl\ to the reader. In the Gaussian approximation we can derive the distribution function of this statistic in analogy to \cref{sec:one-param}. Following \cref{eq:PGst3}, we write
\begin{equation} \label{eq:PGstmulti2}
  \cPGst = \chi^2_{\rm PG,all} - 
  \sum_{S=1}^N d^2_{SP} \,,
\end{equation}
where $\chi^2_{\rm PG,all}$ is the standard PG test statistic for all $N$ data sets plus the prior $P$, and 
\begin{equation}
  d_{SP} = \frac{\widehat{\xi}_S - \widehat{\xi}_P}{\sqrt{s_S + s_P}} \,,\qquad
  S=1,\ldots,N\,.
\end{equation}
Under the null-hypothesis of the model being true, $\chi^2_{\rm PG,all}$ is distributed as $\chi^2_\nu$ with $\nu = 2 N + 1 - 2 = 2N-1$ and the $d_{SP}$ are standard-normal variables with the correlation matrix
\begin{equation}
\rho_{XY}
= \left\{ 
\begin{array}{l@{\qquad}l}
1 & X=Y \,, \\  
\frac{s_P}{\sqrt{(s_X+s_P)(s_Y+s_P)}} & X\neq Y\,,
\end{array} \right.
\label{eq:rhoXY}
\end{equation}
for $X,Y=1,\ldots,N$.

Hence, we write \cref{eq:PGstmulti2} as 
\begin{equation}\label{eq:K-multi}
  \cPGst = \bm{g}^T K\bm{g} \,,
\end{equation}
where $\bm{g}$ is a vector of $\nu = 2N-1$ independent standard normal variables and the $\nu\times\nu$ matrix $K$ is given by 
\begin{equation}
K=I_\nu - \sum_{S=1}^N\bm{a}_S\bm{a}_S^T = I_\nu - AA^T \,,  
\end{equation}
with $\bm{a}_S$ are unit vectors satisfying $\bm{a}_X^T\bm{a}_Y=\rho_{XY}$ and the $\nu\times N$ matrix $A = (\bm{a}_1,\ldots,\bm{a}_N)$. We note that the non-zero eigenvalues of $AA^T$ are the same as the eigenvalues of $A^TA = \rho$. Let us denote these eigenvalues by $\lambda_S^{(\rho)}$, $S=1,\ldots,N$. Then the eigenvalues of $K$ are  
\begin{equation}\label{eq:lambdaK}
  \left(\lambda_i^{(K)}\right) = \big(
\underbrace{1,\ldots,1}_{N-1\ \text{times}},
\quad
1-\lambda_1^{(\rho)},\ldots, 1-\lambda_N^{(\rho)}
\big)
\end{equation}
for $i=1,\ldots,\nu$. Therefore, upon diagonalizing \cref{eq:K-multi} we obtain for the distribution of the test statistic:
\begin{equation}\label{eq:PGstmulti3}
\cPGst
\overset{d}{=}
\sum_{i=1}^\nu \lambda_i^{(K)} G_i^2
\overset{d}{=}
\sum_{i=1}^\nu \lambda_i^{(K)} X_i
\end{equation}
where $G_i$ ($X_i$) are $\nu$ mutually independent standard-normal ($\chi^2_1$) variables. The characteristic function is obtained as
\begin{equation}
  \phi_z(t) = \prod_{i=1}^\nu \left(1 - 2it \lambda_i^{(K)} \right)^{-1/2} \,.
\end{equation}
Hence, in this case the distribution of the test statistic is obtained by calculating the eigenvalues of the correlation matrix \cref{eq:rhoXY}.

\paragraph{Strong prior limit.} For $s_P\to 0$, \cref{eq:rhoXY} gives $\rho \to I_N$ and $\lambda^{(\rho)}_S = 1$ ($S=1,\ldots,N$). In this limit, \cref{eq:lambdaK} for the eigenvalues $\lambda_i^{(K)}$ gives $(N-1)$ times the eigenvalue 1 and all remaining eigenvalues are zero. Therefore, \cref{eq:PGstmulti3} gives a normal $\chi^2$ distribution with $N-1$ degrees of freedom, which is the result of the normal PG for $N$ data sets with one common parameter $\theta$.

\bibliographystyle{JHEP_improved}
\bibliography{./refs}

\end{document}